# Dependence of loss rate of electrons due to elastic gas scattering on the shape of the vacuum chamber in an electron storage ring

**Pradeep Kumar** [a*], **Gurnam Singh** [a], **A.D. Ghodke** [a] **and Pitamber Singh** [b]

[a] *Raja Ramanna Centre for Advanced Technology,*
 *Indore, India*
[b] *Bhabha Atomic Research Centre,*
 *Mumbai, India*
 *E-mail*: `pksharma@rrcat.gov.in`

ABSTRACT: The beam lifetime in an electron storage ring is also limited by the loss rate of the stored electrons due to the elastic coulomb scattering of electrons with the nuclei of residual gas atoms. The contribution to the beam lifetime due to this elastic scattering depends upon the shape factor which is governed by the shape of the vacuum chamber. As the vacuum pressure along the circumference of ring is not uniform so shape factor as a function of longitudinal position is required to be known. In this paper, analytical expressions for the shape factor for a rectangular and an elliptical vacuum chamber as a function of longitudinal position along the circumference in a storage ring are derived using an approach in which the position of electrons at the focusing quadrupole is transformed to the location of defocusing quadrupole and vice versa to define the parts of the vacuum chamber, where the loss of electrons takes place at the location of quadrupoles. The expressions available in the literature are for the average shape factors. The expression of shape factor for a rectangular chamber derived in this paper are similar to the expression for average shape factor quoted in the literature, whereas a new expression for elliptical shape of vacuum chamber having no resemblance with the available expression for average shape factor is obtained. A comparative study of shape factors at each scattering location in the ring obtained from derived expressions and with the existing expressions are reported using Indus-2 lattice parameters for rectangular and elliptical shapes of the chamber. A comparison of average shape factors for these two shapes using derived as well as existing expressions is also reported. These studies indicate that the effect of the rectangular and elliptical shape of vacuum chamber on beam lifetime due to elastic coulomb scattering between electrons and nuclei of residual gas molecules is nearly same.

*Corresponding author

# Contents

## 1. Introduction

For the storage of an electron beam for a longer lifetime in an electron storage ring, average vacuum pressure of the order of 1 nTorr is required in the vacuum chambers in which the electron beam circulates. The presence of residual gas species even at a low pressure of 1 nTorr, causes scattering of the electrons with the nuclei of residual gas atoms. The scattering may be elastic or inelastic. In this paper we discuss the loss of electrons due to the elastic coulomb scattering of electrons with the nuclei of residual gas atoms. In the elastic scattering of electrons of the stored electron beam with the nuclei of residual gas atoms, the electrons are deflected from their path and the amplitude of betatron oscillation of electrons increases. If the amplitude of betatron oscillation of an electron is more than the chamber aperture i.e. the acceptance of the ring at a location, the electron is lost there. The rate of loss of electrons due to this process contributes to beam lifetime. The loss of electrons due to elastic coulomb scattering takes place at minimum acceptance location i.e. at maximum β function in horizontal ($X$) and vertical ($Z$) planes, when the aperture of vacuum chamber is uniform in the ring. The loss of electrons depends on the parameter known as shape factor, which is governed by the shape of the vacuum chamber. In the electron storage rings, rectangular or elliptical shapes of the vacuum chamber are widely used. The vacuum pressure in a storage ring is not uniform along the circumference. It varies from place to place. In order to calculate beam lifetime due to elastic scattering, we have to use the vacuum pressure and shape factor information at all points in the ring. It is thus important to know the shape factor as a function of the longitudinal position in the ring. The average shape factor for a rectangular chamber is discussed in [1] and used for beam lifetime estimation in operating electron storage rings like MAX II [2], SPEAR3 [3], INDUS-2 [4] etc. The average shape factor for elliptical chamber is given in [5]-[6] and used in estimating beam lifetime in SAGA-LS [6] storage ring. In this paper, we derive expressions for the shape factor as a function of longitudinal position along the circumference of a storage ring.

In this paper, general expressions of the shape factor for a rectangular and an elliptical vacuum chamber for a given scattering location in a storage ring are derived starting from ab initio using linear beam dynamics. The motion of electrons is considered to be constrained by the physical aperture neglecting the non linear beam dynamical effects as done by other authors [1]-[6]. Here, the position of electrons at the focusing quadrupole is transformed to the defocusing quadrupole location based on the beta functions in horizontal and vertical planes and vice versa to define the part of the vacuum chamber at which the beam loss takes place at these locations. This approach has enabled derivation of exact expressions for the shape factor as a function of the longitudinal position within the domain of linear beam dynamics. A theoretical estimation of shape factor at different scattering locations in a ring is also carried out using Indus-2 [7]-[8] lattice parameters considering rectangular and elliptical chambers. From the values of shape factor as a function of longitudinal position, the average shape factors are estimated for these shapes and also compared with the average shape factors obtained using the existing expressions.

## 2. Beam lifetime due to elastic coulomb scattering

The loss rate of relativistic electrons due to elastic coulomb scattering $1/\tau_{elastic}$ of electrons with the nuclei of residual gas atom [9] is given by

$$\frac{1}{\tau_{elastic}} = -\frac{1}{N}\frac{dN}{dt} = c\,n\,\sigma$$

where $N, c, n, \sigma$ are the number of electrons, speed of electrons, residual gas density of molecular gas species present in the vacuum chamber and scattering cross section of electrons respectively.

The loss rate of electrons due to elastic coulomb scattering varies from location to location in ring. So

$$\frac{1}{\tau_{elastic}} = \left\langle -\frac{1}{N}\frac{dN}{dt}\right\rangle = c\langle n\sigma\rangle \qquad (1)$$

where $\langle n\sigma\rangle$ is the average of the product of residual gas density and scattering cross section of electrons at different scattering locations spread over the ring.

Differential scattering cross section of elastic coulomb scattering of a relativistic electron scattered by the nuclei of residual gas atom at a location $j$ in ring [9]-[10] is given by

$$\frac{d\sigma_j}{d\Omega} = \frac{Z_i^2\, r_0^2}{4\gamma^2}\frac{1}{\sin^4\left(\frac{\theta}{2}\right)} \qquad (2)$$

where $d\Omega = \sin\theta\, d\theta\, d\phi$ is the solid angle in which the electron is scattered, $\theta$ and $\phi$ are the scattering polar (range from $0$ to $\pi$) and azimuth (range from $0$ to $2\pi$) angle respectively, $Z_i$ is the atomic number of the residual gas atom of species $i$, $r_0$ is the classical electron radius and $\gamma$ is the relativistic Lorentz factor. The electrons will survive in the chamber up to minimum scattering angle $\theta_m$ and will be lost for scattering angle $\theta$ above the value of $\theta_m$. Integrating equation (2) with respect to $\theta$ from limit $\theta_m$ to $\pi$, we get

$$\sigma_j(\phi) = \frac{Z_i^2\, r_0^2}{4\gamma^2}\frac{2\cos^2(\theta_m/2)}{\sin^2(\theta_m/2)} \qquad (3)$$

For small angle of scattering $\theta_m \ll 1$, we get

$$d\sigma_j(\phi) = \frac{2Z_i^2\, r_0^2}{\gamma^2}\frac{d\phi}{\theta_m^2} \qquad (4)$$

The survival chance of a scattered electron depends on the shape and size of the vacuum chamber. As shown in equations (2) and (3), the scattering cross-section has a dependence on angle $\theta$. The maximum allowed displacement for $\theta$ will be different for different shapes and sizes of the vacuum chamber, therefore, the average scattering cross-section as it appears in equation (1), will be different for different shapes and sizes of the vacuum chamber.

The elastic scattering cross section, causing loss of an electron scattered at the location $j$ in a storage ring is given as

$$\sigma_j = \frac{2Z_i^2\, r_0^2}{\gamma^2}F_j \quad \text{where } F_j = \int_0^{2\pi}\frac{d\phi}{\theta_m^2(\phi)} \text{ is defined as the } \textbf{shape factor} \qquad (5)$$

Vacuum pressure or even the composition of residual gas species varies from place to place in a storage ring, therefore it is important to know the shape factor $F$ for each location in a storage ring. If the vacuum pressure and gas composition is uniform along the circumference, the equation (1) becomes

$$\frac{1}{\tau_{elastic}} = c\, n \langle \sigma \rangle \quad \text{where} \quad \langle \sigma \rangle = \frac{2 Z_i^2 r_0^2}{\gamma^2} \langle F \rangle \quad \text{and} \quad \langle F \rangle = \frac{\sum_{j=1}^{l} F_j}{l} \tag{6}$$

where $\langle F \rangle$ is the average shape factor and $l$ is the number of scattering locations spread over the ring at a uniform interval. The loss rate due to elastic scattering $1/\tau_{elastic}$ is proportional to the average shape factor. Obviously, the correct value of average shape factor $\langle F \rangle$ is essential to estimate the beam lifetime due to elastic coulomb scattering between the electrons and the nuclei of residual gas atoms.

## 3. General expressions for shape factor $F$

When an electron collides with a nucleus of a residual gas atom at location $s_0$ in the ring, the electron gets deflected by an angle $\theta$. In spherical polar coordinates, for small scattering angle, the deflection is resolved in $X$ plane as $\theta_x = \theta \cos\phi$ and $\theta_z = \theta \sin\phi$ in $Z$ plane. The electrons are lost at maximum $\beta_x$ or at maximum $\beta_z$ locations as shown in figure 1.

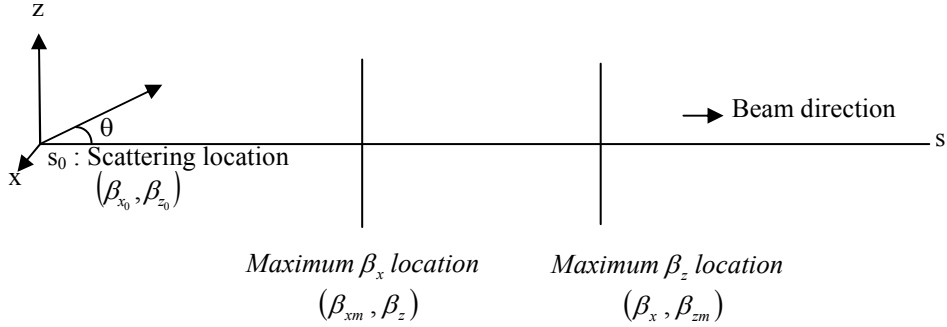

Figure 1. Scattering and beam loss locations in ring

If the electron at location $s_1$ reaches the boundary of the vacuum chamber, its horizontal $x$ and vertical $z$ coordinates at $s_1$ is given as

$$x = \sqrt{\beta_x(s_0)\beta_x(s_1)}\ \theta_x = \sqrt{\beta_x(s_0)\beta_x(s_1)}\ \theta_m \cos\phi \tag{7}$$

$$z = \sqrt{\beta_z(s_0)\beta_z(s_1)}\ \theta_z = \sqrt{\beta_z(s_0)\beta_z(s_1)}\ \theta_m \sin\phi \tag{8}$$

where $\beta_x(s_0)$ and $\beta_z(s_0)$ are $\beta$ functions at the scattering location $s_0$ in $X$ and $Z$ planes respectively and $\beta_x(s_1)$ and $\beta_z(s_1)$ are $\beta$ functions at beam loss location $s_1$ in $X$ and $Z$ planes respectively and $\theta_m$ is the minimum scattering angle. In equation (7) and (8) as well as everywhere in this paper, maximum betatron displacements are considered taking the betatron phase term equal to one because here only those electrons, which are lost from the ring, are taken into account. Substituting $x$ and $z$ from equation (7) and (8) into equation (5), the expression for shape factor $F_j$ at location $j$ in ring becomes

$$F_j = \int_0^{2\pi} \frac{d\phi}{\theta_m^2(\phi)} = \int_0^{2\pi} \frac{\beta_x(s_0)\beta_x(s_1)\cos^2\phi + \beta_z(s_0)\beta_z(s_1)\sin^2\phi}{x^2 + z^2} d\phi$$

$$\text{where } \tan\phi = \sqrt{\frac{\beta_x(s_0)\beta_x(s_1)}{\beta_z(s_0)\beta_z(s_1)}} \frac{z}{x}$$

(9)

Dependence of the shape factor $F_j$ on $x$ and $z$ indicates that it is governed by the shape of the vacuum chamber. We consider that the storage ring has a vacuum chamber of uniform cross section all along the circumference. In such a ring, the beam loss will takes place either at the location where $\beta_x$ is maximum or the location where $\beta_z$ is maximum. In the following paragraph, we derive expressions for the shape factors for a rectangular and an elliptical vacuum chamber.

### 3.1 Rectangular vacuum chamber

Let the horizontal and vertical dimensions of the rectangular vacuum chamber be $a$ and $b$ respectively. In order to find out the domain of azimuth angle $\phi$ for electron loss in $X$ and $Z$ plane, we assume that at maximum $\beta_z$ location, the electrons lie on the boundary of vacuum chamber. The electron, which, is at $P(a,b)$ at maximum $\beta_z$ location, will be at $P''\left(\sqrt{\beta_{xm}/\beta_x}\,a, \sqrt{\beta_z/\beta_{zm}}\,b\right)$ at maximum $\beta_x$ location as shown in figure 2(a), where $\beta_{xm}$ and $\beta_z$ are $\beta$ functions at maximum $\beta_x$ location in $X$ and $Z$ planes respectively, $\beta_x$ and $\beta_{zm}$ are $\beta$ functions at maximum $\beta_z$ location in $X$ and $Z$ planes respectively. The electrons which lie on the boundary of the chamber at maximum $\beta_z$ will, accordingly, follow the dotted rectangle at maximum $\beta_x$ location as shown in figure 2(a) whereas the solid rectangle shows the actual aperture at maximum $\beta_x$ location.

Similarly, we assume that at maximum $\beta_x$ location, electrons are on the boundary of the vacuum chamber. Referring to figure 2(b), the electron, which is at $P(a,b)$ at maximum $\beta_x$ location will be at $P''\left(\sqrt{\beta_x/\beta_{xm}}\,a, \sqrt{\beta_{zm}/\beta_z}\,b\right)$ at maximum $\beta_z$ location. The electrons which lie on the boundary of the chamber at maximum $\beta_x$ will follow the dotted rectangle at maximum $\beta_z$ location as shown in figure 2(b) whereas the solid rectangle shows the actual aperture at maximum $\beta_z$ location.

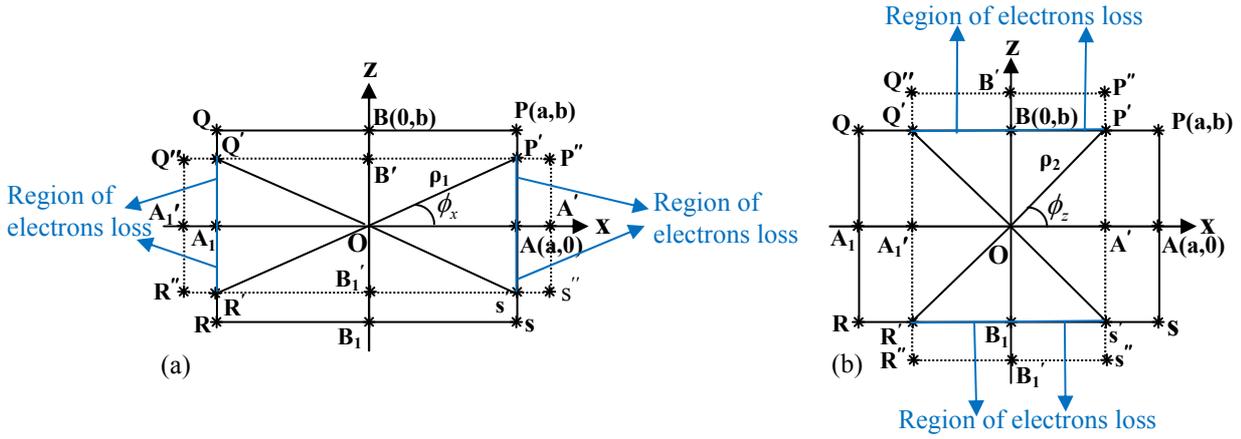

*Figure 2. Electron Positions (a) at maximum $\beta_x$ (dotted rectangle) with respect to their positions on the boundary of chamber at maximum $\beta_z$ and (b) at maximum $\beta_z$ (dotted rectangle) with respect to their positions on the boundary of chamber at maximum $\beta_x$*

From figure 2(a), it is seen that at the location of maximum $\beta_x$, electrons, which have the magnitude of horizontal displacement $a$ or greater and also vertical displacement magnitude up to $OB'\left(\sqrt{\beta_z/\beta_{zm}}\,b\right)$ are lost on the $P'AS'$ and $Q'A_1R'$ parts of the vacuum chamber. Electrons, having the magnitude of vertical displacement greater than to $OB'$ are lost at the maximum $\beta_z$ location.

Similarly, from figure 2(b), it is understood that at the location of maximum $\beta_z$, electrons, which have the magnitude of vertical displacement $b$ or greater and also magnitude of horizontal displacement up to $OA'\left(\sqrt{\beta_x/\beta_{xm}}\,a\right)$ are lost on the $P'BQ'$ and $S'B_1R'$ parts of the vacuum chamber. Electrons having the magnitude of horizontal displacement greater than to $OA'$ are lost at the maximum $\beta_x$ location.

In order to obtain the shape factor, we consider only the first quadrant of the vacuum chamber taking advantage of four fold symmetry. First, we consider the maximum $\beta_x$ location and for this we refer to figure 2(a). Here, the electrons are lost on the aperture boundary $P'A$. The azimuth angle $\phi$ varies from 0 to an angle $\phi_{xm}$ which is related to the angle $\phi_x$ at this location. Similarly at the maximum $\beta_z$ location, electrons are lost on boundary $P'B$ of figure 2(b). The azimuth angle $\phi$ here varies from $\phi_{zm}$ to $\pi/2$ which is related to the angle $\phi_z$ at this location.

From figure 2(a), the coordinate of the electron at the location $P'$ is $\left(a,\sqrt{\beta_z/\beta_{zm}}\,b\right)$ where $\beta_{zm} > \beta_z$. From equation (7) and (8), the coordinate of $P'(x,z)$ at maximum $\beta_x$ location is given as

$$x = \sqrt{\beta_{x_0}\beta_{xm}}\,\theta_m\cos\phi$$

$$z = \sqrt{\beta_{z_0} \beta_z}\ \theta_m \sin\phi$$

$$\tan\phi = \sqrt{\frac{\beta_{x_0} \beta_{xm}}{\beta_{z_0} \beta_z}}\ \frac{z}{x} \tag{10}$$

The maximum value of $\phi$ i.e. $\phi_{xm}$ is obtained at location $P'$ of electron loss, $x = a$ and $z = \sqrt{\beta_z/\beta_{zm}}\ b$, so

$$\tan\phi_{xm} = \sqrt{\frac{\beta_{x_0}\beta_{xm}}{\beta_{z_0}\beta_z}}\sqrt{\frac{\beta_z}{\beta_{zm}}}\frac{b}{a} \Rightarrow \tan\phi_{xm} = \sqrt{\frac{\beta_{x_0}\beta_{xm}}{\beta_{z_0}\beta_{zm}}}\frac{b}{a} \Rightarrow \phi_{xm} = \tan^{-1}\left(\frac{pb}{a}\right)$$

$$\text{where } p = \sqrt{\frac{\beta_{x_0}\beta_{xm}}{\beta_{z_0}\beta_{zm}}} \tag{11}$$

From figure 2(a), $\tan\phi_x = \dfrac{z}{x} \Rightarrow \tan\phi = p_1 \tan\phi_x$ where $p_1 = \sqrt{\dfrac{\beta_{x_0}\beta_{xm}}{\beta_{z_0}\beta_z}}$ (using eq. 10) (12)

Similarly from figure 2(b), the coordinate of the electron at point $P'$ is $\left(\sqrt{\beta_x/\beta_{xm}}\ a,\ b\right)$, where $\beta_{xm} > \beta_x$. The coordinate of $P'(x,z)$ at maximum $\beta_z$ location is given as

$$x = \sqrt{\beta_{x_0}\beta_x}\ \theta_m \cos\phi$$

$$z = \sqrt{\beta_{z_0}\beta_{zm}}\ \theta_m \sin\phi$$

$$\tan\phi = \sqrt{\frac{\beta_{x_0}\beta_x}{\beta_{z_0}\beta_{zm}}}\ \frac{z}{x}$$

Using the coordinate of $P'$, $x = \sqrt{\dfrac{\beta_x}{\beta_{xm}}}\ a$ and $z = b$, the maximum value of $\phi$ i.e. $\phi_{zm}$ is

$$\tan\phi_{zm} = \sqrt{\frac{\beta_{x_0}\beta_x}{\beta_{z_0}\beta_{zm}}}\sqrt{\frac{\beta_{xm}}{\beta_x}}\frac{b}{a} \Rightarrow \tan\phi_{zm} = \sqrt{\frac{\beta_{x_0}\beta_{xm}}{\beta_{z_0}\beta_{zm}}}\frac{b}{a} \Rightarrow \phi_{zm} = \tan^{-1}\left(\frac{pb}{a}\right)$$

$$\text{where } p = \sqrt{\frac{\beta_{x_0}\beta_{xm}}{\beta_{z_0}\beta_{zm}}} \tag{13}$$

From figure 2(b), $\tan\phi_z = \dfrac{z}{x} \Rightarrow \tan\phi = p_2 \tan\phi_z$ where $p_2 = \sqrt{\dfrac{\beta_{x_0}\beta_x}{\beta_{z_0}\beta_{zm}}}$ (14)

It is clear from equations (11) and (13) that $\phi_{xm} = \phi_{zm} = \phi_m$. In brief, for $\phi$ varying from 0 to $\phi_m$, the electrons are lost at maximum $\beta_x$ location on $P'A$ (figure 2(a)) and from $\phi_m$ to $\pi/2$ at maximum $\beta_z$ on $P'B$ (figure 2(b)).

The shape factor $F_j$ taking into consideration the four fold symmetry of chamber is

$$F_j = \int_0^{2\pi} \frac{d\phi}{\theta_m^2(\phi)} = 4\int_0^{\pi/2} \frac{d\phi}{\theta_m^2(\phi)}$$

In this integral, in first quadrant, electrons scattered into azimuth angle $\phi$ between 0 to $\phi_m$ will be lost at maximum $\beta_x$ and those scattered between $\phi_m$ to $\pi/2$ will be lost at maximum $\beta_z$ location.

$$\int_0^{\pi/2} \frac{d\phi}{\theta_m^2(\phi)} = \int_0^{\phi_m} \frac{\beta_{x_0}\beta_{xm}\cos^2\phi + \beta_{z_0}\beta_z\sin^2\phi}{\rho_1^2} d\phi + \int_{\phi_m}^{\pi/2} \frac{\beta_{x_0}\beta_x\cos^2\phi + \beta_{z_0}\beta_{zm}\sin^2\phi}{\rho_2^2} d\phi$$

From figure 2(a), $\rho_1^2 = a^2 + a^2\tan^2\phi_x$, putting in first integral and from figure 2(b), $\rho_2^2 = b^2 + b^2\cot^2\phi_z$, putting in second integral and using relation (12) and (14), we get

$$F_j = \frac{2\beta_{x_0}\beta_{xm}}{a^2}\left[\phi_m + \frac{1}{2}\sin 2\phi_m\right] + \frac{2\beta_{z_0}\beta_{zm}}{b^2}\left[\frac{\pi}{2} - \phi_m + \frac{1}{2}\sin 2\phi_m\right]$$

$$F_j = \frac{2\beta_{x_0}\beta_{xm}}{a^2}\left[\tan^{-1}\left(\frac{pb}{a}\right) + \frac{pab}{a^2 + p^2b^2}\right] + \frac{2\beta_{z_0}\beta_{zm}}{b^2}\left[\cot^{-1}\left(\frac{pb}{a}\right) + \frac{pab}{a^2 + p^2b^2}\right]$$

$$\text{where } p = \sqrt{\frac{\beta_{x_0}\beta_{xm}}{\beta_{z_0}\beta_{zm}}} \tag{15}$$

The above expression, which gives the contribution to the shape factor due to the elastic coulomb scattering at location $j$ is similar to that given for the average shape factor in [1] as below

$$\langle F\rangle = \frac{2\langle\beta_x\rangle\beta_{xm}}{a^2}\left[\tan^{-1}\left(\frac{pb}{a}\right) + \frac{pab}{a^2 + p^2b^2}\right] + \frac{2\langle\beta_z\rangle\beta_{zm}}{b^2}\left[\cot^{-1}\left(\frac{pb}{a}\right) + \frac{pab}{a^2 + p^2b^2}\right]$$

$$\text{where } p = \sqrt{\frac{\langle\beta_x\rangle\beta_{xm}}{\langle\beta_z\rangle\beta_{zm}}}, \langle\beta_x\rangle \text{ and } \langle\beta_z\rangle \text{ are average values of } \beta \text{ function in the ring} \tag{16}$$

The expression (16) is derived in [1] by assuming the average value of minimum scattering angle $\theta_x$ and $\theta_z$ avoiding point to point calculation of shape factor in the ring which we have considered in deriving expression (15) by using above approach. In section 4, it is seen that the average shape factor estimated from expression (15) using equation (6) and average shape factor obtained from expression (16) are nearly same.

### 3.2 Elliptical vacuum chamber

We follow the same approach as that used for a rectangular chamber to find out an expression for the shape factor of an elliptical chamber. Here also we assume that at maximum $\beta_z$ location, electrons are on the boundary of chamber surface. The position of these electrons at maximum $\beta_x$ location will be as shown in figure 3(a) by the dotted ellipse, whereas the solid ellipse is the boundary of vacuum chamber surface at maximum $\beta_x$ location. Similarly we assume that the electrons at maximum $\beta_x$ location are on the boundary of vacuum chamber, these electrons at maximum $\beta_z$ location will lie on the dotted ellipse as shown in figure 3(b) whereas the solid ellipse here shows the boundary of the vacuum chamber at maximum $\beta_z$ location.

From figure 3(a), it is clear that at maximum $\beta_x$ location, electrons are lost on the $PAS$ and $QCR$ parts of the vacuum chamber. Similarly figure 3(b) indicates that at maximum $\beta_z$ location, electrons are lost on $QBP$ and $RDS$ parts of the vacuum chamber.

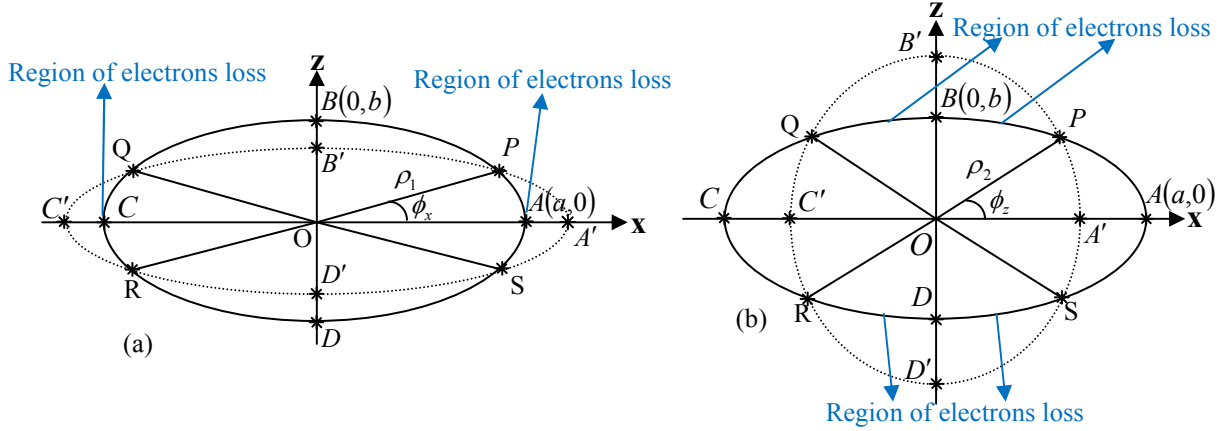

*Figure 3. Electron positions (a) at maximum $\beta_x$ (dotted ellipse) with respect to their positions on the boundary of chamber at maximum $\beta_z$ and (b) at maximum $\beta_z$ (dotted ellipse) with respect to their positions on the boundary of chamber at maximum $\beta_x$*

From Figure 3(a), equation of solid ellipse representing the vacuum chamber is $\dfrac{x^2}{a^2}+\dfrac{z^2}{b^2}=1$ and that of the dotted ellipse is $\dfrac{x^2}{a_1^2}+\dfrac{z^2}{b_1^2}=1$, where $a_1=\sqrt{\dfrac{\beta_{xm}}{\beta_x}}\,a$ and $b_1=\sqrt{\dfrac{\beta_z}{\beta_{zm}}}\,b$. Let $\beta_{x_0}$ and $\beta_{z_0}$ be the $\beta$ functions at the location $s_0$ where scattering of electron with gas atom takes place. At location $s_0$, the electron gets kick of angle $\theta_m$, the position coordinates $P(x,z)$ at the location of maximum $\beta_x$ is given as

$$x=\sqrt{\beta_{x_0}\,\beta_{xm}}\;\theta_m\,\cos\phi$$
$$z=\sqrt{\beta_{z_0}\,\beta_z}\;\theta_m\,\sin\phi$$

From this equation, we get $\tan\phi=\sqrt{\dfrac{\beta_{x_0}\,\beta_{xm}}{\beta_{z_0}\,\beta_z}}\,\dfrac{z}{x}$ (17)

In figure 3(a), point P is intersection of two ellipses so on solving these, we get

$$\frac{z}{x} = \sqrt{\frac{\beta_z (\beta_{xm} - \beta_x)}{\beta_{xm} (\beta_{zm} - \beta_z)}} \frac{b}{a}$$

Accordingly from equation (17), the maximum value of $\phi$ at the location of electron loss $\phi_{xm}$ is,

$$\tan \phi_{xm} = \sqrt{\frac{\beta_{x_0} (\beta_{xm} - \beta_x)}{\beta_{z_0} (\beta_{zm} - \beta_z)}} \frac{b}{a} \Rightarrow \phi_{xm} = \tan^{-1}\left(\frac{pb}{a}\right) \text{ where } p = \sqrt{\frac{\beta_{x_0} (\beta_{xm} - \beta_x)}{\beta_{z_0} (\beta_{zm} - \beta_z)}} \qquad (18)$$

Let the coordinate of point P on ellipse in figure 3(a) be $(\rho_1 \cos\phi_x, \rho_1 \sin\phi_x)$ so, we get

$$\frac{1}{\rho_1^2} = \frac{\cos^2\phi_x}{a^2} + \frac{\sin^2\phi_x}{b^2}, \quad \tan\phi_x = \frac{z}{x} \Rightarrow \tan\phi = p_1 \tan\phi_x \text{ where } p_1 = \sqrt{\frac{\beta_{x_0} \beta_{xm}}{\beta_{z_0} \beta_z}} \quad \text{(using eq. 17)} \quad (19)$$

Similarly from figure 3(b), equation of solid ellipse representing the vacuum chamber is $\frac{x^2}{a^2} + \frac{z^2}{b^2} = 1$ and that of dotted ellipse is $\frac{x^2}{c_1^2} + \frac{z^2}{d_1^2} = 1$, where $c_1 = \sqrt{\frac{\beta_x}{\beta_{xm}}} a$ and $d_1 = \sqrt{\frac{\beta_{zm}}{\beta_z}} b$.

The position coordinates $P(x,z)$ at the location of maximum $\beta_z$ is given as

$$x = \sqrt{\beta_{x_0} \beta_x} \, \theta_m \cos\phi$$
$$z = \sqrt{\beta_{z_0} \beta_{zm}} \, \theta_m \sin\phi$$

In figure 3(b), point P is intersection of two ellipses, so on solving these, we get

$$\frac{z}{x} = \sqrt{\frac{\beta_{zm} (\beta_{xm} - \beta_x)}{\beta_x (\beta_{zm} - \beta_z)}} \frac{b}{a}$$

The maximum value of $\phi$ at the location of electron loss $\phi_{zm}$ is given as

$$\tan \phi_{zm} = \sqrt{\frac{\beta_{x_0} (\beta_{xm} - \beta_x)}{\beta_{z_0} (\beta_{zm} - \beta_z)}} \frac{b}{a} \Rightarrow \phi_{zm} = \tan^{-1}\left(\frac{pb}{a}\right) \text{ where } p = \sqrt{\frac{\beta_{x_0} (\beta_{xm} - \beta_x)}{\beta_{z_0} (\beta_{zm} - \beta_z)}} \qquad (20)$$

It is clear from equations (18) and (20) that $\phi_{xm} = \phi_{zm} = \phi_m$. In brief for $\phi$ varying from 0 to $\phi_m$, the electrons are lost at maximum $\beta_x$ location and from $\phi_m$ to $\pi/2$, they are lost at the maximum $\beta_z$ location.

Let the coordinate of point P on ellipse in figure 3(b) be $(\rho_2 \cos\phi_z, \rho_2 \sin\phi_z)$ so

$$\frac{1}{\rho_2^2} = \frac{\cos^2\phi_z}{a^2} + \frac{\sin^2\phi_z}{b^2}, \quad \tan\phi_z = \frac{z}{x} \Rightarrow \tan\phi = p_2 \tan\phi_z, \text{ where } p_2 = \sqrt{\frac{\beta_{x_0} \beta_x}{\beta_{z_0} \beta_{zm}}} \qquad (21)$$

Shape factor $F_j$ is given as

$$F_j = \int_0^{2\pi} \frac{d\phi}{\theta_m^2(\phi)} = 4 \int_0^{\pi/2} \frac{d\phi}{\theta_m^2(\phi)}$$

$$\int_0^{\pi/2} \frac{d\phi}{\theta_m^2(\phi)} = \int_0^{\phi_m} \frac{\beta_{x_0}\beta_{xm}\cos^2\phi + \beta_{z_0}\beta_z \sin^2\phi}{\rho_1^2} d\phi + \int_{\phi_m}^{\pi/2} \frac{\beta_{x_0}\beta_x \cos^2\phi + \beta_{z_0}\beta_{zm}\sin^2\phi}{\rho_2^2} d\phi$$

Using $\rho_1^2$ from equation (19) and $\rho_2^2$ from equation (21), we get

$$\int_0^{\pi/2} \frac{d\phi}{\theta_m^2(\phi)} = \int_0^{\phi_m} \left(\beta_{x_0}\beta_{xm}\cos^2\phi + \beta_{z_0}\beta_z \sin^2\phi\right)\left(\frac{\cos^2\phi_x}{a^2} + \frac{\sin^2\phi_x}{b^2}\right) d\phi +$$

$$\int_{\phi_m}^{\pi/2} \left(\beta_{x_0}\beta_x \cos^2\phi + \beta_{z_0}\beta_{zm}\sin^2\phi\right)\left(\frac{\cos^2\phi_z}{a^2} + \frac{\sin^2\phi_z}{b^2}\right) d\phi$$

Using equation (19), $\tan\phi = p_1 \tan\phi_x$ in first integral and equation (21), $\tan\phi = p_2 \tan\phi_z$ in second integral, we get

$$F_j = \frac{2\beta_{x_0}\beta_{xm}}{a^2}\left(\phi_m + \frac{1}{2}\sin 2\phi_m\right) + \frac{2\beta_{z_0}\beta_z}{b^2}\left(\phi_m - \frac{1}{2}\sin 2\phi_m\right) +$$

$$\frac{2\beta_{z_0}\beta_{zm}}{b^2}\left(\frac{\pi}{2} - \phi_m + \frac{1}{2}\sin 2\phi_m\right) + \frac{2\beta_{x_0}\beta_x}{a^2}\left(\frac{\pi}{2} - \phi_m - \frac{1}{2}\sin 2\phi_m\right) \qquad (22)$$

$$\text{where } \phi_m = \tan^{-1}\left(\sqrt{\frac{\beta_{x_0}(\beta_{xm} - \beta_x)}{\beta_{z_0}(\beta_{zm} - \beta_z)}} \frac{b}{a}\right)$$

$$F_j = \frac{2\beta_{x_0}\beta_{xm}}{a^2}\left(\tan^{-1}\left(\frac{pb}{a}\right) + \frac{pab}{a^2 + p^2 b^2}\right) + \frac{2\beta_{x_0}\beta_x}{a^2}\left(\cot^{-1}\left(\frac{pb}{a}\right) - \frac{pab}{a^2 + p^2 b^2}\right) +$$

$$\frac{2\beta_{z_0}\beta_{zm}}{b^2}\left(\cot^{-1}\left(\frac{pb}{a}\right) + \frac{pab}{a^2 + p^2 b^2}\right) + \frac{2\beta_{z_0}\beta_z}{b^2}\left(\tan^{-1}\left(\frac{pb}{a}\right) - \frac{pab}{a^2 + p^2 b^2}\right) \qquad (23)$$

$$\text{where } p = \left(\sqrt{\frac{\beta_{x_0}(\beta_{xm} - \beta_x)}{\beta_{z_0}(\beta_{zm} - \beta_z)}}\right)$$

This is a new expression of the shape factor for an elliptical shape of vacuum chamber which has not been reported so far.

An expression of average shape factor for an elliptical chamber is given in [5]-[6] as below.

$$\langle F \rangle = \pi \left[\frac{\langle\beta_x\rangle \beta_{xm}}{a^2} + \frac{\langle\beta_z\rangle \beta_{zm}}{b^2}\right] \qquad (24)$$

where $\langle\beta_x\rangle$ and $\langle\beta_z\rangle$ are the average $\beta$ function in $X$ and $Z$ planes, $\beta_{xm}$ and $\beta_{zm}$ are the maximum $\beta$ function in $X$ and $Z$ planes respectively. This expression can be derived considering the loss of electrons at the location where $\beta_x$ as well as $\beta_z$ are maximum (APPENDIX-A). Such situations are unlikely to occur in modern storage rings. While deriving the expression (24), the expression (A.5) representing dependence of shape factor on longitudinal position $s$ is obtained and it is used later in this paper for comparison of shape factors of two shapes.

## 4. Estimation of point by point shape factor and its average in Indus-2 ring

Indus-2 is an electron storage ring which is operational at beam energy 2.5 GeV beam energy at RRCAT, Indore, India [7]-[8]. The point by point shape factor $F_j$ was estimated by using Indus-2 lattice parameter for rectangular and elliptical shape of the vacuum chamber. The lattice functions $\beta_x$ and $\beta_z$ in one unit cell of Indus-2 at the interval of 5 cm was estimated using computer code MAD [11] which are shown in figure 4.

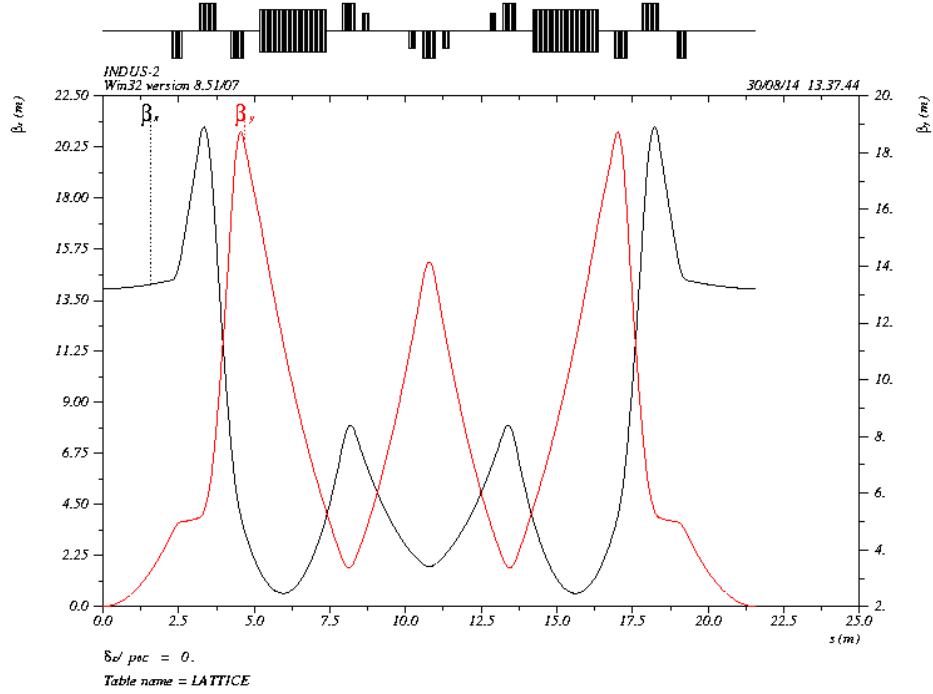

*Figure 4. $\beta$ functions in one unit cell of Indus-2 electron storage ring*

### 4.1 Comparison of shape factor with longitudinal position for rectangular and elliptical chamber

The shape factor at each scattering location was estimated for rectangular and elliptical chamber using derived expression (15) for rectangular chamber and expression (23) for elliptical chamber using Indus-2 lattice parameters $\beta_{xm} = 21.1\,m$, $\beta_{zm} = 18.7\,m$, $\beta_x = 3.9\,m$, $\beta_z = 5.4\,m$, $\langle\beta_x\rangle = 8.1\,m$ and $\langle\beta_z\rangle = 7.9\,m$. The variation of shape factors with longitudinal position $s$ in one unit cell of Indus-2 for rectangular $(a = 30\,mm, b = 15\,mm)$ and elliptical $(a = 30\,mm, b = 15\,mm)$ chamber shapes are shown in figure 5(a). The increase in shape factor at each scattering location from rectangular to elliptical shape of chamber is shown in figure 5(b).

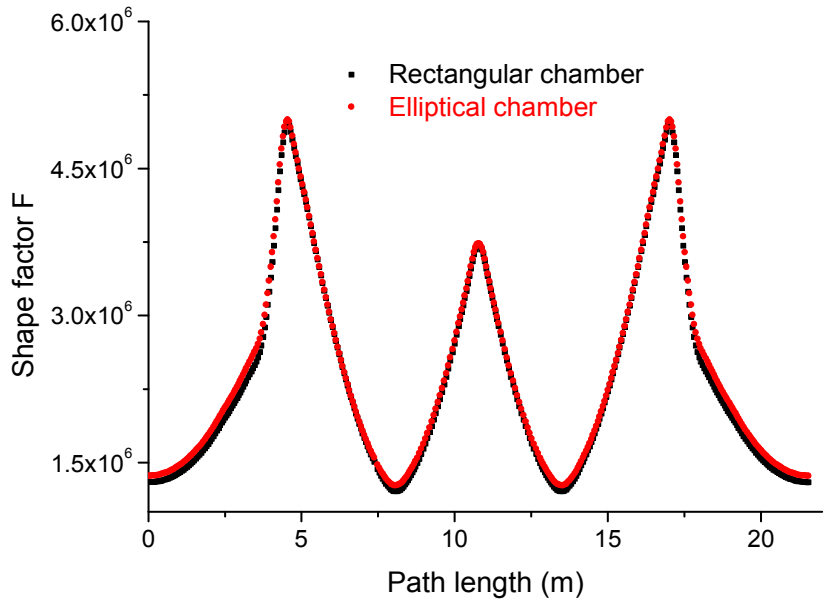

*Figure 5(a). Comparison of shape factor in one unit cell for rectangular and elliptical shape*

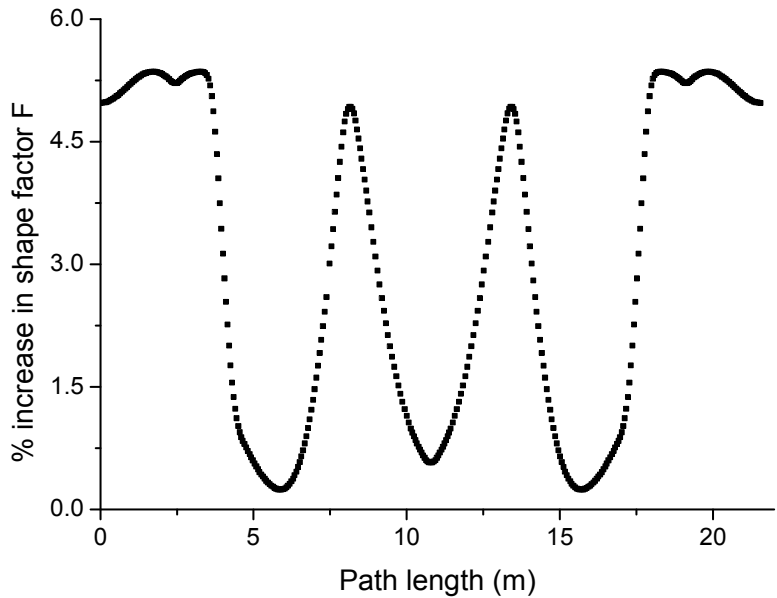

*Figure 5(b). Increase in shape factor at each scattering location from rectangular to elliptical shape*

As we have seen that we got a similar expression for rectangular chamber but a new expression was obtained for elliptical chamber. A comparison in shape factor at each scattering location using expression (15) for rectangular chamber and existing expression (A.5) for elliptical chamber of the same dimension as above is shown in figure 6(a) and the increase in shape factor from rectangular to elliptical shape is shown in figure 6(b).

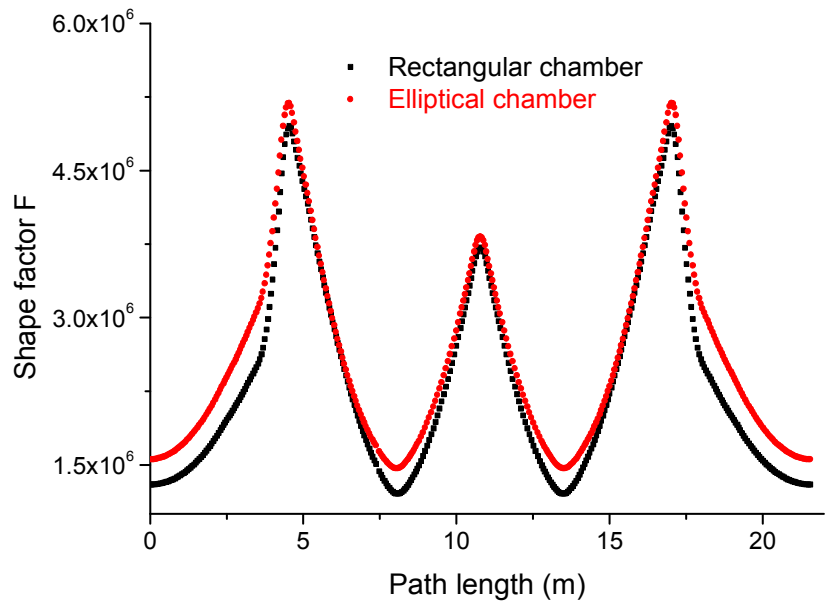

Figure 6(a). Comparison of shape factor in one unit cell for rectangular and elliptical shape

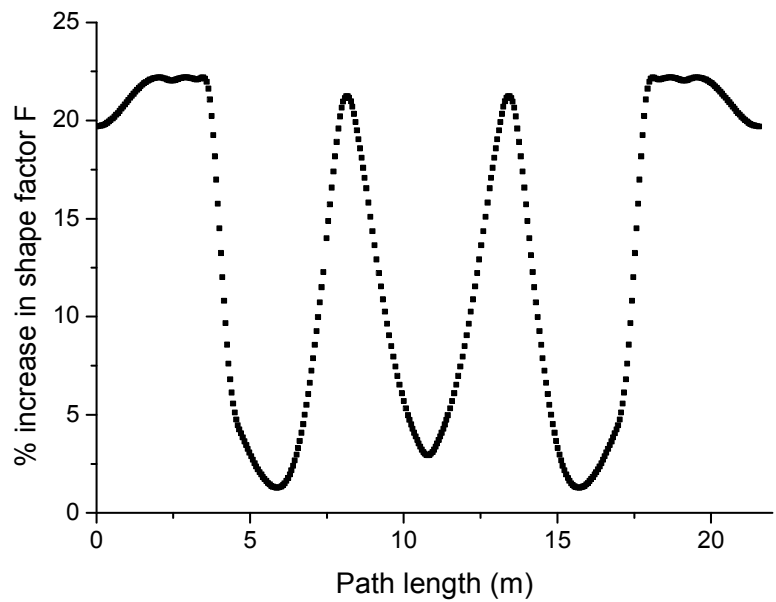

*Figure 6(b). Increase in shape factor at each scattering location from rectangular to elliptical shape*

Comparison of figure 5(b) with figure 6(b) indicates that the difference between the shape factors for two shapes as per derived expressions is much less than that as perceived with existing expressions.

## 4.2 Comparison of average shape factor for rectangular and elliptical shape

From the value of shape factor at different scattering locations, average shape factor was estimated for rectangular $(a=30mm, b=15mm)$ and elliptical $(a=30mm, b=15mm)$ chamber using equation (6) and comparison with the values obtained using existing expression (16) and (24) for rectangular and elliptical shape respectively are given in Table 1.

*Table 1. Comparison of estimated average shape factor using derived and existing expressions for rectangular and elliptical shape of chamber of dimension $a = 30\,mm$ and $b = 15\,mm$.*

| Chamber shape | Average shape factor using equation (6) and expressions (15) and (23) ($F_1 \times 10^6$) | Average shape factor using existing expressions (16) and (24) ($F_2 \times 10^6$) |
| --- | --- | --- |
| Rectangular | 2.35 | 2.31 |
| Elliptical | 2.41 | 2.66 |
| *Percentage increase in average shape factor F* | *2.5%* | *15.1%* |

## 4.3 Comparison of shape factor with longitudinal position for square and circular chamber

From the expression (15) of shape factor for rectangular shape, putting dimensions $a$ and $b$ equal, we get shape factor for square shape and in expression (23), putting $a=b$, we get the shape factor for circular shape of the vacuum chamber. The comparison of shape factor at each scattering location in one unit cell of Indus-2 for square $(a=b=30\,mm)$ and circular $(radius=30\,mm)$ shape of chamber is shown in figure 7(a) and the percentage increase in shape factor at each scattering location is shown in figure 7(b).

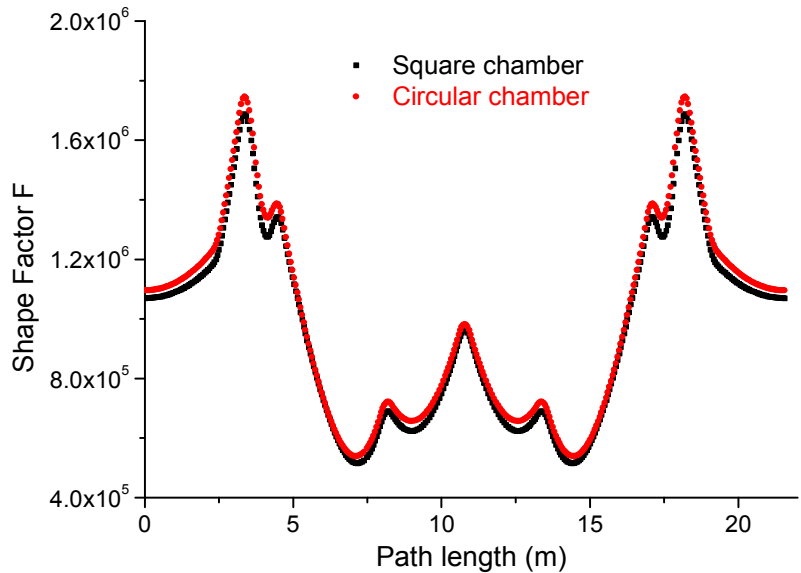

*Figure 7(a). Comparison of shape factor in one unit cell for square and circular shape*

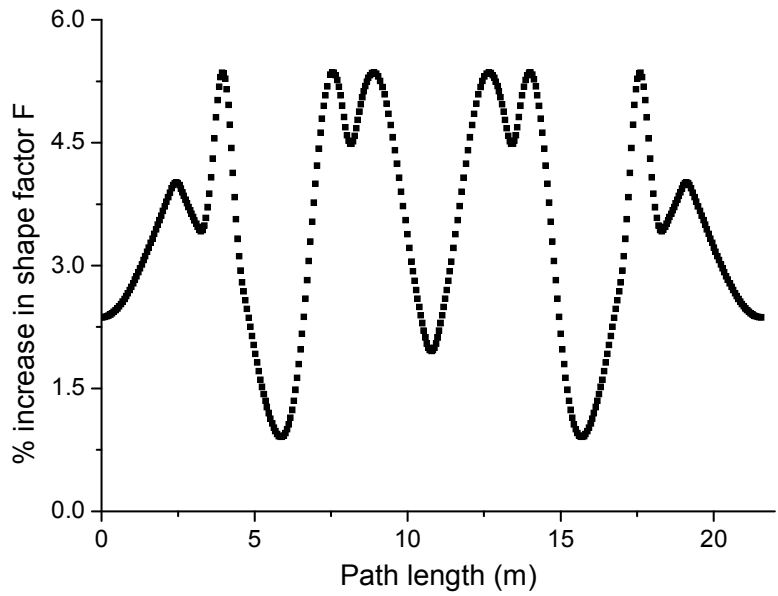

*Figure 7(b).Increase in shape factor at each scattering location from square to circular shape*

A comparison in shape factor at each scattering location using expression (15) for square ($a=b=30\,mm$) and existing expression (A.5) for circular ($radius=30\,mm$) chamber is shown in figure 8(a) and the increase in shape factor from square to circular shape of vacuum chamber is shown in figure 8(b).

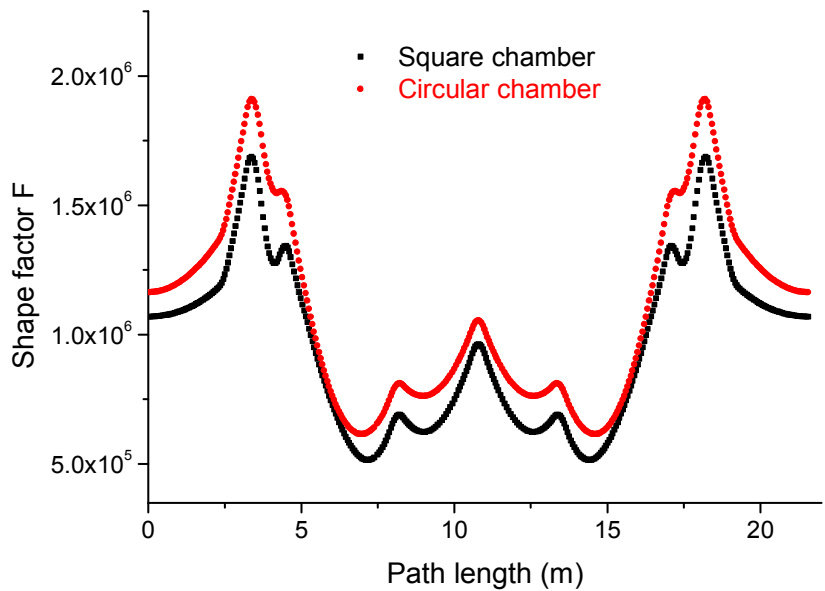

*Figure 8(a). Comparison of shape factor in one unit cell for square and circular shape*

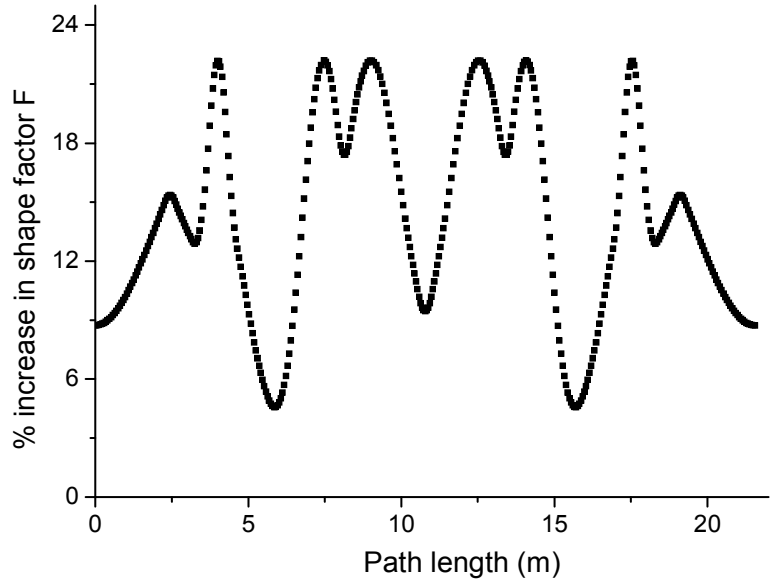

*Figure 8(b). Increase in shape factor at each scattering location from square to circular shape*

Comparison of figure 7(b) with figure 8(b) reveals that the difference between the shape factors for two shapes as per derived expressions is much less than that obtained with existing expressions.

### 4.4 Comparison of average shape factor for square and circular shape

From the value of shape factor at different scattering locations, average shape factor was estimated for square ($a=b=30\,mm$) and circular ($radius=30\,mm$) chamber using equation (6) and comparison with the values obtained using existing expressions is given in Table 2.

*Table 2. Comparison of estimated average shape factor using derived and existing expressions for square ($a=b=30\,mm$) and circular ($radius=30\,mm$) shape of the chamber.*

| Chamber shape | Average shape factor using equation (6) and expressions (15) and (23) ($F_1 \times 10^5$) | Average shape factor using existing expressions (16) and (24) ($F_2 \times 10^5$) |
|---|---|---|
| Square | 9.63 | 9.11 |
| Circular | 9.95 | 11.12 |
| *Percentage increase in average shape factor F* | *3.3%* | *22.0%* |

From table 1, if we compare the change in average shape factor from rectangular to elliptical shape ($a=30\,mm, b=15\,mm$), the increase in shape factor from the derived expressions is ~2.5% whereas it increases to ~15.1% if we use the existing expressions. Similarly as reported in [1], the average shape factor increase from square to circular shape of chamber is ~22% as seen in table 2, whereas from the derived expressions, the increase is ~3.3%. The large difference between the shape factor values of rectangular and elliptical chambers or higher values of shape factors for elliptical shapes as per the existing expressions are attributed to approximations made in the integration of equation (9) in [5]-[ 6].

## 5. Conclusion

The general expressions of shape factor due to elastic coulomb scattering of electrons with the nuclei of residual gas atoms at a scattering location in an electron storage ring are derived for a rectangular chamber using linear beam dynamics and using the same approach, shape factor for an elliptical chamber is also derived. The method of transforming the position of electrons at the focusing quadrupole to the defocusing quadrupole location and vice versa has made it possible to derive exact analytical expressions for the shape factor for the two shapes of the vacuum chamber. The difference between the shape factors of the rectangular and elliptical chambers obtained from these expressions are much less than that obtained using existing expressions as the geometrical effects of shapes are taken into account as required. The expression obtained for the shape factor at a given location for rectangular shape is similar to the expression for the average shape factor available in the literature and the average shape factors using derived and existing expression are nearly same. The expression derived for the shape factor for an elliptical shape is new and has no similarity to the expression available in the literature because the approach followed in this paper takes into account of loss of electrons in the elliptical shape in the way as it happens according to the linear beam dynamics and not at one location, where both $\beta_x$ and $\beta_z$ are maximum. The expressions for shape factor as a function of longitudinal position derived in this paper are required for estimation of the beam lifetime due to elastic scattering of electrons with nuclei of residual gas atoms when the vacuum pressure in the ring is not the same everywhere along the circumference as normally observed in storage rings.

# APPENDIX-A

Assuming that the loss of electrons takes place at one location. This implies that both $\beta_x$ and $\beta_z$ have maxima at this point, which is rarely the case for example, it can occur in weak focusing storage rings. Then we may substitute the following $x$ and $z$ in equation 9

$$x = \sqrt{\beta_{x_0} \beta_{xm}}\; \theta_m \cos\phi \qquad \text{A.1}$$

$$z = \sqrt{\beta_{z_0} \beta_{zm}}\; \theta_m \sin\phi \qquad \text{A.2}$$

we get

$$F_j = \int_0^{2\pi} \frac{d\phi}{\theta_m^2(\phi)} = \int_0^{2\pi} \frac{\beta_x(s_0)\beta_{xm}\cos^2\phi + \beta_z(s_0)\beta_{zm}\sin^2\phi}{x^2 + z^2} d\phi \qquad \text{A.3}$$

At the location of electron loss, $x^2 + z^2 = \rho^2$ and $(\rho\cos\phi, \rho\sin\phi)$ are arbitrary coordinates of $P$. The coordinate $P(\rho\cos\phi, \rho\sin\phi)$ lies on ellipse $\dfrac{x^2}{a^2} + \dfrac{z^2}{b^2} = 1$, so using

$$\frac{1}{\rho^2} = \frac{\cos^2\phi}{a^2} + \frac{\sin^2\phi}{b^2}\; \text{in equation A.3 we get}$$

$$F_j = \int_0^{2\pi} \left[\beta_x(s_0)\beta_{xm}\cos^2\phi + \beta_z(s_0)\beta_{zm}\sin^2\phi\right] \left[\frac{\cos^2\phi}{a^2} + \frac{\sin^2\phi}{b^2}\right] d\phi \qquad \text{A.4}$$

After integration, we get similar expression as 24

$$F_j = \pi \left[\frac{\beta_x(s_0)\beta_{xm}}{a^2} + \frac{\beta_z(s_0)\beta_{zm}}{b^2}\right] \qquad \text{A.5}$$

This is the expression of shape factor at location $j$ in ring for elliptical shape of vacuum chamber. Taking average $\beta$ in ring, we get average shape factor $\langle F \rangle$ similar to expression 24

$$\langle F \rangle = \pi \left[\frac{\langle \beta_x \rangle \beta_{xm}}{a^2} + \frac{\langle \beta_z \rangle \beta_{zm}}{b^2}\right] \qquad 24$$

In deriving this equation, we have assumed that the loss takes place at one location where both $\beta_x$ and $\beta_z$ are maximum.

## Acknowledgement

Authors are thankful to P.D. Gupta, Director, RRCAT and P.R. Hannurkar, Head, IOAPDD for continuous encouragement, support and guidance to pursue this work. One of the authors (GS) is thankful to Department of Atomic Energy for the award of Raja Ramanna Fellowship.

# References

[1] H. Wiedemann, *Particle Accelerator Physics*, Third edition, Springer-Verlag, Berlin, (2007).

[2] Erik Wallen, *Aperture and lifetime measurements with movable scrapers at MAX II, Nuclear Instruments and Methods in Physics Research A* 508 (2003).

[3] X. Huang and J. Corbett, *Measurement of beam lifetime and applications for SPEAR3, Nuclear Instruments and Methods in Physics Research A* 629 (2011).

[4] Pradeep Kumar et al., *Measurements of aperture and beam lifetime using movable beam scrapers in Indus-2 electron storage ring, Review of Scientific Instruments,* 84,123301 (2013).

[5] Jim Murphy, *Synchrotron light source data book*, Brookhaven National Laboratory, 42333, May (1996).

[6] T. Kaneyasu, Y. Takabayashi, Y. Iwasaki, S. Koda, *Beam lifetime study based on momentum acceptance restriction by movable beam scraper*, Nuclear Instruments and Methods in Physics Research A 694 (2012).

[7] G. Singh, G.K. Sahoo, D. Angal, B. Singh, A.D. Ghodke and P. Kant, *Synchrotron radiation source Indus-2*, *Indian Journal of Applied Physics*, 35(1) (1997).

[8] *Technical report on design of synchrotron radiation source Indus-2, Raja Ramanna Centre for Advanced Technology, Indore (India)*, September (1998).

[9] A. Wrulich, *Single beam lifetime*, in proceedings of CERN accelerator school, Editor: S. Turner, Fifth general accelerator physics course vol. I, CERN 94-01 (1994).

[10] C. Bocchetta, *Lifetime and beam quality*, in proceedings of CERN accelerator school on synchrotron radiation and free electrons lasers, Editor: S. Turner*, CERN 98-04 (1998).

[11] F. Christoph Iselin, "*The MAD Program: Physical Methods Manual*", CERN/SL/92 (AP), (1994).